\documentstyle[11pt,a4wide,epsf]{article}

\newcommand{\pspicture}[1]{
\centerline{\setlength\epsfxsize{9.2cm}\epsfbox{#1}}}

\newcommand{\ga}{~\raisebox{-2pt}{$\stackrel{>}{\sim}$}~}

\newcommand{\be}{\begin{equation}}
\newcommand{\ee}{\end{equation}}
\newcommand{\bea}{\begin{eqnarray}}
\newcommand{\eea}{\end{eqnarray}}
\newcommand{\fr}{\frac}
\def \s{\sigma}
\def \n{\hfil\break}

\begin{document}
\pagestyle{plain}
\begin{flushright}
DAMTP Preprint: 97-110\\
\end{flushright}
\vskip 20pt

\begin{center}
{\huge \bf Heavy Hybrids from NRQCD} \\[15pt]
{\sc UKQCD Collaboration} \\[10pt]
T. Manke \footnote{T.Manke@damtp.cam.ac.uk},
I.T. Drummond\footnote{itd@damtp.cam.ac.uk},
R.R. Horgan\footnote{rrh@damtp.cam.ac.uk} ,
H.P. Shanahan\footnote{H.P.Shanahan@damtp.cam.ac.uk}\\[10pt]
{\sl DAMTP, University of Cambridge, Cambridge CB3 9EW, England}
\end{center}

\begin{abstract}
We present a quenched lattice calculation for the lowest lying $b \bar
b g$-hybrid states
in the framework of NRQCD using the leading order
Hamiltonian up to ${\cal O}(mv^2)$.
We demonstrate the existence of a nearly degenerate rotational
band of states with an excitation energy approximately 1.6 GeV above
the $\Upsilon$ ground state. This lies around the $B \bar B_J^*$-threshold
but well above the $B \bar B$-threshold. 
Therefore a heavy hybrid signal may well be detected if the
centre-of-mass energy in
B-factories is raised a few hundred MeV to coincide with other
resonances above the 4S state.
Our prediction is consistent with most phenomenological models and 
lattice calculations carried out in the static limit.
\end{abstract}

\section{Introduction}

Hybrid mesons are of intense interest both theoretically
and experimentally because of the opportunity they provide for
investigating nonperturbatively the gluonic degrees of freedom in QCD.
In contrast to the standard $q\bar q$-mesons in which the quarks form 
a colour singlet, hybrids contain quarks in a colour octet state.
There exist hybrid states that have quantum numbers not available to
pure $q\bar q$-states and as a result do not mix with them. These {\em exotic}
states are of particular interest. Interest has been further
heightened by the recently reported discovery at Brookhaven of
a $1^{-+}$ state at (1370  $\pm16$ ${+50}\atop{-30}$) ${\rm MeV}$
\cite{brook}.
However, the experimental study of such light hybrids is made
difficult by the density of levels in the 1-2 GeV range and by strong
mixing effects. Such difficulties are minimised for heavy hybrids
and this should result in a clearer signal at the appropriate energies
in future B-factories.

The study of hybrid states has been approached in a number of ways,
such as flux tube models \cite{flux_tube}, bag models
\cite{bag_hybrid}, sum rules \cite{sum_rules} and the constituent gluon model
\cite{constituent_gluon}.  Light hybrids have been 
studied in the framework of lattice QCD \cite{cm_light,milc}. 
Heavy hybrids have been approached through the static quark limit
\cite{cm_and_peran}
by identifying a $q\bar q$-potential appropriate to each
gluonic excitation \cite{ford}. 

In this paper we study hybrid excitations of the $b\bar b$-system
using NRQCD \cite{nrqcd_improved}. This allows us to go beyond the 
static limit.
Our calculation is preliminary in that we use only the lowest
order heavy quark Hamiltonian and employ the quenched approximation
for the gluon field.
Previous work confirms that 
this approximation gives a reasonable account of low lying spin averaged 
energy levels of heavy quark systems \cite{nrqcd_precision,omv6}. 
In this sense we believe we have achieved an acceptable computation
of the lowest {\it magnetic} hybrid excitations. 
We have not yet computed the hybrid levels with opposite parity.
Within bag models these {\it electric} hybrids are believed to 
lie higher in energy than the corresponding magnetic states \cite{bag_hybrid}.

In Section 2 we introduce the non-relativistic evolution equation for
the quark propagator. 
In Section 3 we discuss continuum operators that connect the hybrid 
states to the vacuum and in Section 4 we set out the lattice
versions of these operators that were used in our simulation. 
The results are presented in Section 5.

\section{NRQCD and Heavy Quark Propagators}
The NRQCD approach to the computation of the heavy quark propagator
$G(x,y)$, has been explained previously \cite{nrqcd_precision,omv6}. For completeness we record here 
the Euclidean time evolution equation, namely 
\be
G({\bf x},t+1;y)=\left(1-\frac{aH_0}{2n}\right)^n U_t^{\dag}(x)
\left(1-\frac{aH_0}{2n}\right)^n G({\bf x},t;y) ~~,~~~~~t \ge t_y ~~.
\label{eq:evol}
\ee
The initial condition has the form
\be
G({\bf x},t=t_y;y)=S({\bf x,y}) ~~.
\label{initcond}
\ee
where $S({\bf x,y})$ is the source term on the first timeslice, ($t=t_y$),
appropriate to the channel under study.
In this paper the Hamiltonian is
\be
H_0 = -  \frac{\Delta^2}{2m_b}~~ ,
\ee
where $m_b$ is the bare quark mass and $\Delta^2$ is the standard
spatial covariant Laplacian defined in \cite{nrqcd_improved}.
In the calculation all the link variables
are tadpole improved according to the replacement $U_\mu(x)\rightarrow U_\mu(x)/u_0$
where $u_0$ is obtained from the plaquette $U_{\Box}$: $u_0=\langle 0|\frac{1}{3}{\rm Tr}U_{\Box}|0 \rangle^{1/4}$. Other suggested
improvements \cite{trottier}, make no significant difference to the present calculation
which is concerned only with spin averaged quantities. This is consistent
with our neglect of higher corrections to the Hamiltonian that incorporate
quark spin-gluon couplings and ${\cal O}(mv^4)$ terms.

\section{Hybrid States and Operators}
We will use the standard nomenclature for hybrid states \cite{hybrid_constituent}
in which charge conjugation and parity satisfy
\be
C=(-1)^{l+s+1}
\ee
\be
P=\left\{\begin{array}{ll}(-1)^{l+j}&\mbox{TE}\\(-1)^{l+j+1}&\mbox{TM}\end{array}\right.~~,
\ee
where $l$ and $s$ are the $q\bar q$ orbital angular momentum and spin, and $j$ is 
the gluon angular momentum. The historical notations TE and TM refer to the magnetic and
electric hybrid states, respectively. 

To extract masses we calculate two-point functions of operators
with the appropriate quantum numbers. For hybrid states these operators must
include a gluon field factor in order to ensure the presence of a gluon excitation.
This is done by introducing the gauge field operators
\bea
B_i&=&\fr{1}{2}\epsilon_{ijk}F_{jk}~~,\\ \nonumber
E_i&=&F_{it}~~,
\eea
where $F_{\mu\nu}$ is the gluon field tensor. We can alternatively
replace {\bf E} by ${\bf \nabla \times B}$ to access the same quantum
numbers on a single timeslice.

From $B_i$ we can construct the $j=1$ {\it magnetic} hybrid operators. The spin singlet
states are coupled to the vacuum by the operators defined in Table \ref{tab:mag_hybrid_singlet}.
\begin{table}[ht]
\begin{center}
\begin{tabular}{|c|c|r|}\hline
state&$l$&$J^{PC}$\\ \hline\hline
$\chi^{\dag}B_i\psi$&0&$1^{--}$\\ \hline
$\chi^{\dag}\{B_i, D_i\}\psi$&1&$0^{++}$\\ \hline
$\chi^{\dag}\epsilon_{ijk}\{B_j, D_k\}\psi$&1&$1^{++}$\\ \hline
$\chi^{\dag}(\{B_i, D_j\}+\{B_j, D_i\}-\fr{1}{3}\delta_{ij}\{B_i, D_j\})\psi$&1&$2^{++}$\\ \hline
\end{tabular}
\end{center}
\caption{Continuum operators for spin-singlet hybrid states. The
non-relativistic 2 spinors, $\psi$ and $\chi^{\dag}$, denote the quark
field and anti-quark field, respectively.}
\label{tab:mag_hybrid_singlet}
\end{table}

It has been pointed out by Griffiths {\it et al.} \cite{griffiths} that there is a common $q\bar q$-potential interaction for all 
the $j=1$ states in the static quark limit. The presence of excited glue, however, results
in a relatively shallow structure for the radial dependence of the potential
function. Consequently we can anticipate that the orbital motion of {\it finite} mass quarks 
in the presence of this potential will give rise to 
a nearly degenerate rotational band of states. In particular
the $l=0$ and $l=1$ states will be approximately degenerate. If, in addition, the orbital dynamics
of the quarks is controlled by a spin-independent force, as is the case in the 
approximation used in this paper, the triplet quark states built on the above
singlet states will also lie in the same degenerate band.
The triplet states shown in Table \ref{tab:mag_hybrid_triplet} are
built on the $1^{--}$ state and contain the $1^{-+}$
which has exotic quantum numbers. 
\begin{table}[ht]
\begin{center}
\begin{tabular}{|c|c|r|}\hline
state&$l$&$J^PC$\\ \hline\hline
$\chi^{\dag}\s_iB_i\psi$&0&$0^{-+}$\\ \hline
$\epsilon_{ijk}\chi^{\dag}\s_jB_k\psi$&0&$1^{-+}$\\ \hline
$\chi^{\dag}(\s_iB_j+\s_jB_i-\fr{1}{3}\delta_{ij}\s_iB_i)\psi$&0&$2^{-+}$\\ \hline
\end{tabular}
\end{center}
\caption{Continuum operators for spin-triplet states corresponding to
the $1^{--}$.}
\label{tab:mag_hybrid_triplet}
\end{table}
Triplet states built on the $1^{++}$ are shown in Table
\ref{tab:mag_hybrid_triplet_zero}.
Those contain also the exotics $0^{+-}$ and $2^{+-}$.
\begin{table}[ht]
\begin{center}
\begin{tabular}{|c|c|r|}\hline
state&$l$&$J^PC$\\ \hline\hline
$\chi^{\dag}\s_i\epsilon_{ijk}\{B_j, D_k\}\psi$&1&$0^{+-}$\\ \hline
$\chi^{\dag}\s_j(\{B_i, D_j\}-\{B_j, D_i\})\psi$&1&$1^{+-}$\\ \hline
$\chi^{\dag}(\s_i\epsilon_{jkl}\{B_k, D_l\}+\s_j\epsilon_{ikl}\{B_k, D_l\} 
-\fr{1}{3}\delta_{ij}\s_i\epsilon_{jkl}\{B_k, D_l\})\psi$&1&$2^{+-}$\\ \hline
\end{tabular}
\end{center}
\caption{Continuum operators for spin-triplet states corresponding to
the $1^{++}$.}
\label{tab:mag_hybrid_triplet_zero}
\end{table}
There are further triplet states including a $3^{+-}$ state. All these 
states can be expected to be essentially degenerate in the case of our 
simple spin-independent Hamiltonian.
\n\n

\section{Lattice Operators}
Because our evolution equation involves no spin-corrections to the
Hamiltonian we focus on the lattice versions of the operators in Table
\ref{tab:mag_hybrid_singlet}. In constructing them we replace covariant
derivatives, $D_i$, with covariant lattice derivatives, $\Delta_i^n$, in the extended form
used in \cite{omv6,ron_hybrid}. For the magnetic gluon operators which involve
only spatial derivatives this presents no problem since these
operators can be formulated in terms of variables on a single
timeslice. 
We define the colour magnetic field on the lattice as
\be
B_i^n=\frac{1}{2}\epsilon_{ijk}[\Delta_j^n,\Delta_k^n]~~,
\ee
with the definition

\bea
\Delta_i^n \psi(x) &\equiv& L_i^n \psi(x+ni) -  L_{-i}^n
\psi(x-ni)\nonumber \\
L_i^n \psi(x) &\equiv& U_i(x)U_i(x+i) \ldots U_i(x+(n-1)i) \psi(x+ni)
\eea

As in an earlier study \cite{ron_hybrid}, these operators
result in commutators of extended link variables for the hybrid state

\be
B_i^n=\frac{1}{2}\epsilon_{ijk}\left\{[L^n_j,L^n_k]-[L^n_{-j},L^n_{-k}]\right\}~~,
\ee

For the free field case, where $U_\mu(x)=1$, these operators vanish as
expected.
Finally we have the extended hybrid operator
\be
H_i^n({\bf x}) = \epsilon_{ijk}\chi^{\dag}({\bf x})([L_j^n,L_k^n] -
[L_{-j}^n,L_{-k}^n])\psi({\bf x})~~.
\ee

{\renewcommand{\arraystretch}{1.2}
\begin{table}[ht]
\begin{center}
{\small
\begin{tabular}{|c|c|c|}
\hline
$O_h$ irrep. & $\chi^{\dag}(x)~O~\Psi(x)$  & lowest contiuum $J^{PC}$  \\
\hline
$A_1$ & 1          & $0^{-+}$  \\
\hline
$T_1$ & $\epsilon_{ijk}~\Delta_j\Delta_k$    & $1^{--}$  \\
\hline
$A_1$ & $\epsilon_{ijk}\Delta_i\Delta_j\Delta_k$     & $0^{++}$ \\
$T_1$ & $\epsilon_{ijk}~\epsilon_{klm}\{\Delta_j,\Delta_l\Delta_m\}$  & $1^{++}$ \\
$T_2$ & $s_{ijk}\epsilon_{klm}\{\Delta_j,\Delta_l\Delta_m\}$  & $2^{++}$  \\
$E$   & $S_{\alpha jk}\epsilon_{klm}\{\Delta_j,\Delta_l\Delta_m\}$  & $2^{++}$  \\
\hline
\end{tabular}}
\caption{Equivalent lattice operators of Table \ref{tab:mag_hybrid_singlet}. The first column
denotes the irreducible representation of the octaheadral group. We
define $s_{ijk} = |\epsilon_{ijk}|$ and $S_{\alpha jk}$ projects out
the two linearly independent traceless-symmetric combinations
corresponding to the representation E.}
\end{center}
\label{tab:trafo}
\end{table}
}

We have achieved a significant improvement for the signal by employing
the fuzzing algorithm for link variables suggested in
\cite{cm_fuzz}. We use a central link weight of $c=2.5$ with
six fuzzing iterations.
We now use those {\it fuzzed} link variables to construct the extended
links, which we then use in the meson operators. 
Phenomenological models indicate that hybrid states are more extended
than the standard $q \bar q$-states. This suggests that operators with
large spatial separation will have the best overlaps. In compromising
between spatial size of the operators and the efficiency of the code
we choose n=4 and 5.
In some cases the operators were further
improved by using, in addition, Jacobi-smearing for the quark fields
\cite{jacobi_smearing}.

The meson correlator is written as a Monte Carlo average over all configurations
\be
{\rm C}^{nm}(x,y) = \langle {\bf tr}\left[G^{\dag}(x,y)G^{nm}(x,y)\right] \rangle~~,
\label{eq:correlator}
\ee
where ${\bf tr}$ denotes contraction over all internal degrees
of freedom and $G^{nm}$ is the smeared propagator defined by 
\be
G^{nm}(x,y) \equiv \sum_{z_1,z_2} O^n(x,z_1)G(z_1,z_2)O^{m \dag}(z_2,y)~~.
\ee
Here $(n,m)$ stands for the radii at (sink, source) and
the $O^n$ are operators as defined in Table \ref{tab:trafo}.
For the extended propagator we solve equation \ref{eq:evol} with
$S({\bf x,y}) = O^{m\dag}({\bf x,y})$ and multiply with $O^n$ at the sink.
We fix the origin at some (arbitrary) lattice
point, $y$, and sum over all spatial ${\bf x}$ so
as to project out the zero momentum mode. 

\section{Simulation and Results}
The parameters of our simulation are shown in Table \ref{tab:parameters}.
\begin{table}
\begin{center}
\begin{tabular}{|l|l|}
\hline
$\beta$          & 6.0 \\
$L^3 \times T$  & $16^3 \times 48$ \\
$a^{-1}$ in GeV (from $1P-1S$) & 2.44(4) \\
$am_b$, $n$ & 1.71, 2 \\
$u_0$       & 0.878  \\
spatial starts & 8   \\
temporal starts& 5    \\
total number of measurements & 20, 000  \\
\hline
\end{tabular}
\caption{The parameters in our simulation. 
To increase the statistics, we chose several starting points per configuration.}
\label{tab:parameters}
\end{center}
\end{table}
The quenched gauge field configurations were all generated at the EPCC
in Edinburgh. The propagators were calculated at the HPCF in Cambridge.

\begin{table}
\begin{center}
\begin{tabular}{|l|l|l|l|l|l|}
\hline
Operator & State & \multicolumn{2}{|c|}{Fit results} &  \multicolumn{2}{|c|}{Energies [GeV]} \\
         &       & $aE_0$  & $aE_1-aE_0$   & $E_0-E_{\Upsilon}$  & $E_1-E_0$ \\
\hline
\hline
1 & $0^{-+}$ & 0.4487(13) & -- & 0. &  --\\
\hline
\hline
$\sum_i B_i$ & $1^{--}$ & 1.114(40) & 0.628(66)  & 1.62(10) & 1.53(16) \\
\hline
\hline
$\Delta_i\cdot B_i$ &  $0^{++}$   & 1.175(22)& 0.609(73)  &
1.775(61) & 1.49(18)  \\
$\{\Delta_{[i}, B_{j]}\} \to \sum_i [\Delta_i,\Delta^2]$ &  $1^{++}$
& 1.134(19)& 0.61(10)   & 1.674(54) & 1.49(25)  \\
$\{\Delta_{\{1}, B_{2\}}\}$ & $2^{++}(T_2)$  & 1.161(19)& 0.600(61)
& 1.740(55) & 1.47(15)     \\
$\{\Delta_1, B_1\} - \{\Delta_2, B_2\}$ &  $2^{++}(E)$ & 1.126(13)&
0.520(58)  & 1.655(42) & 1.27(14) \\
\hline
\end{tabular}
\caption{Hybrid spectrum from 499 configurations at $\beta=6.0$, $am_b =
1.71$, $u_0=0.878$, $a_{1P-1S}^{-1}=2.44(4)$ GeV. 
$E_0$ and $E_1$ denote the energies of ground and first excited states
in the relevant channel. The results are obtained
from a two-exponential fit as shown in column 3 and 4.
In column 5 and 6 we have converted all results into physical units, using the
inverse lattice spacing which is determined from the $1P-1S$ splitting.
The ground state in each channel is given relative to the $\Upsilon(1S)$.}
\label{tab:hybrid_results}
\end{center}
\end{table}

\begin{figure}
\pspicture{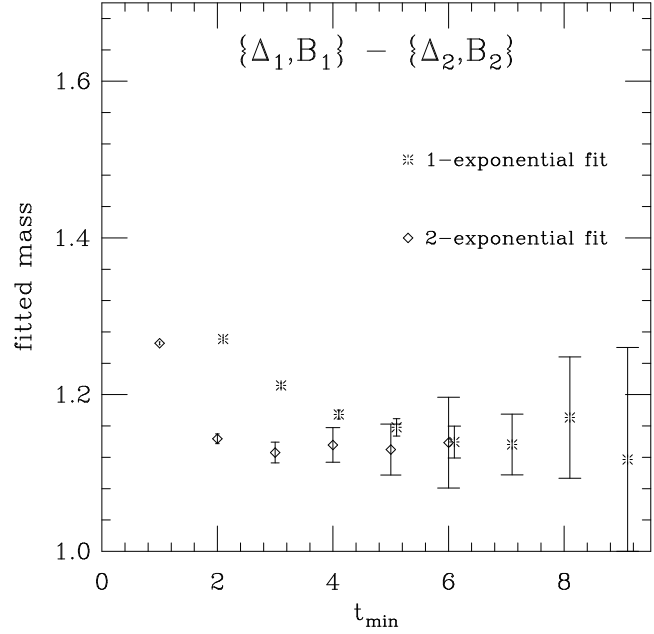}
\caption{Fit to hybrid signal. Consistent fit results are shown for
the ground state for both a 1-exponential fit and a 2-exponential fit.}
\label{fig:hybrid_tmin}
\end{figure}

We fit the correlators to the multi-exponential form
\be
C_\alpha^{nm}(t) = \sum_{i=1}^{\rm n_{fit}} a_{\alpha i}^{nm} e^{-M_i^{\alpha} t}~~.
\label{eq:theory}
\ee
Here $\alpha$ denotes a meson state with certain quantum numbers, $(n,m)$ the
different radii at the (sink, source) and $t$ the Euclidean time. 
As the hybrid excitations are very high above the ground state the
signal-to-noise ratio deteriorates very quickly and vanishes at $t \ga 15$.
Therefore we choose $t_{\rm max} = 12$ and vary $t_{\rm min}$ in search of a plateau.
An example of such a plot for the lowest lying $b \bar b$-hybrid is
shown in Figure \ref{fig:hybrid_tmin}.
A collection of our results can be found in Table
\ref{tab:hybrid_results} and Figure \ref{fig:hybrid_results}.

We quote the masses from plateaus where $\chi^2/{\rm dof} < 1$. 
The error estimate is calculated from the inverse of
the Hessian matrix which is determined during a Levenberg-Marquardt
$\chi^2$ minimisation \cite{numrec}.

\section{Conclusions}
\begin{figure}
\pspicture{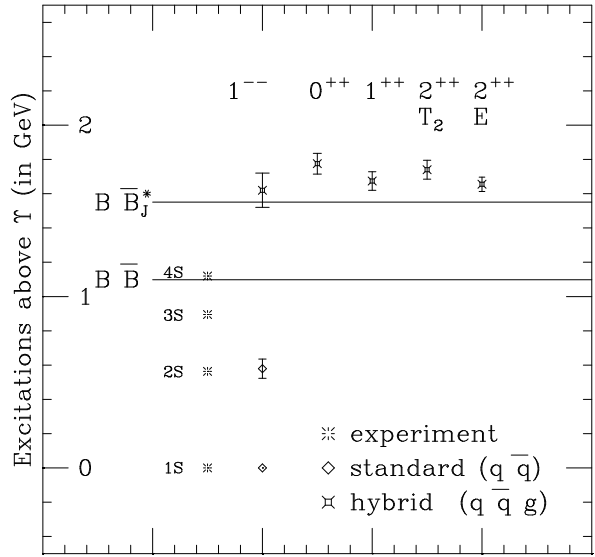}
\caption{Results. The $1^{--}$ is 1.62(10) GeV above the 1S. The
corresponding rotational band appears to be degenerate around the
$B\bar B_J^{\ast}$-threshold.
Each state is the spin-averaged and some of them contain also the
exotics $1^{-+}, 0^{+-}, 2^{+-}$. See main text.}
\label{fig:hybrid_results}
\end{figure}

In this paper we reported on a heavy hybrid signal obtained within the
framework of NRQCD. 
This field theoretical calculation goes beyond the static
approximation and so has the virtue of incorporating the dynamics
of the heavy quarks.
In our investigation we calculated the masses of hybrids where the $q
\bar q$-pair is in a colour octet coupled to the colour magnetic
field. 
Some of the results have already been reported elsewhere
\cite{our_hybrid}. Work on the same configurations using different
operators has been discussed in \cite{sara_hybrid}.
We obtained a signal for several correlation functions where
the magnetic field couples to spin-singlet states of different orbital
angular momenta, as listed in Table \ref{tab:trafo}.
Static potential models predict that the above states should lie in a
nearly degenerate rotational band.
From Table \ref{tab:hybrid_results}~and Figure \ref{fig:hybrid_results}~
it can be seen that our results confirm this picture to within two
standard deviations.
For our Hamiltonian all the spin-triplet states are degenerate in
energy with the spin-singlets and we are only sensitive to
spin-averaged quantities. Therefore, the exotic $1^{-+}$ state is
degenerate in this approximation with the $1^{--}$ and likewise both
the $0^{+-}$ and $2^{+-}$ are degenerate with the $1^{++}$.

Phenomenologically, there are two important thresholds for the production
and decay of hybrid $b \bar b$-states. These are the  $B \bar
B$-threshold at 10.56 GeV and the  $B \bar B_J^{\ast}$-threshold at 11.01 GeV.
For the latter we assume that $B_{J}^{\ast}(5732)$ is indeed a P-wave
as suggested in the Particle Data Book \cite{pdb}. Sometimes this
state is also refered to as $B^{\ast\ast}$. 
Below the (S+P)-threshold hybrid states are thought to be stable
\cite{close_break}. 
As can be seen from Figure \ref{fig:hybrid_results} our results
suggest that hybrid states lie close to the (S+P)-threshold and 4-5
standard deviations above the (S+S)-threshold.

In our calculation there are a number of sources of systematic errors.  
First, we have retained only terms ${\cal O}(mv^2)$ in the Hamiltonian and 
neglected those ${\cal O}(mv^4)$ and higher. Numerically this
may not be a bad approximation since we expect the quarks in the
hybrid states to be even more non-relativistic than in the 
$\Upsilon$ itself. This is consistent with the very shallow
$q \bar q$-potential predicted in the static limit of the hybrid
state \cite{cm_and_peran,kuti_hybrid,colin_hybrid} and the implied near
degeneracy of the resulting rotational band 
- a result confirmed in our calculation. However a study of singlet-triplet
splitting requires that higher order terms, such as
${\bf \sigma \cdot B}$, be taken into account. 
The above argument suggests that this splitting is small.
Furthermore, there may be finite size effects because the hybrids in the
shallow potential are expected to be larger than the $\Upsilon$. From
studies in the static limit one expects the interquark seperation in
$b\bar bg$ to be of the order of 0.5 fm
\cite{sara_hybrid,colin_hybrid}. 
The actual extent of a hybrid state may still be  bigger than this
\cite{bag_hybrid}. 
We note that for the parameters used in this paper the lattice has a
spatial extent of approximately 1.3 fm.

Finally, and perhaps most importantly, we must take into account
uncertainties in the value of $a^{-1}$, the inverse lattice spacing.
These uncertainties are intrinsic to any calculation based on a quenched
gluon approximation, since not all mass ratios can be simultaneously correct.
The result therefore depends on which observables are used to fit $a^{-1}$~. In
the numbers quoted above we used the $1P-1S$ mass difference of a standard
NRQCD calculation of the $\Upsilon $-system. This is a consistent approach
and yields $a^{-1}=2.44(4)$ GeV and our number of 11.08(10) GeV for the lowest hybrid. 
Perantonis and Michael quote 10.81(25) GeV 
from a Schr\"odinger Equation using the static potential from the
lattice, where $a_\sigma^{-1}$ is determined from the string tension in the quenched
approximation \cite{cm_and_peran}.
Their error also includes an estimate for quenching effects.
If we use their value of $a_\sigma^{-1}$=2.04(2) GeV, we obtain the mass of the
$1^{-+}$ to be 10.82(8) GeV, which is consistent with their result. 
These uncertainties should be resolved in a simulation with dynamical
light quarks where the coupling runs appropriately and
the inverse lattice spacing is expected to be the same when determined at
different scales.
Therefore the issue whether the hybrid states lie above or below the
$B \bar B^\ast$-threshold, must be addressed using
unquenched gauge field configurations. 
Subject to this uncertainty
in the value of the inverse lattice spacing, our results are consistent with
the predictions from most phenomenological models
\cite{flux_hybrid}.

\begin{center}{\bf Acknowledgements}\end{center}
We would like to thank other members of the UKQCD collaboration in
particular S. Collins for useful discussions.
T.M. is supported by a grant from EPSRC (Ref. No. 94007885).
H.P.S. is supported by the Leverhulme Trust.
Our calculations are performed on  the Hitachi SR2001 located at the
University of Cambridge High Performance Computing Facility and the
CRAY-T3D at the Edinburgh Parallel Computing Centre at Edinburgh University.


\begin{thebibliography}{99}
\bibitem{brook} D.R. Thompson {\it et al.} Phys. Rev. Lett. 79, (1997), 1630-1633.
\bibitem{flux_tube} N. Isgur and J. Paton, Phys. Rev. {\bf D}31, (1985), 2910-2929.
\bibitem{bag_hybrid} P. Hasenfratz {\it et al.}, Phys. Lett. {\bf B}95,
(1981), 299-305. \\
T. Barnes {\it et al.}, Nucl. Phys. {\bf B}224, (1983), 241-264.
\bibitem{sum_rules} J. Govaerts {\it et al.}, Nucl. Phys. {\bf B}284, (1987), 674-689.
\bibitem{constituent_gluon} Yu.S. Kalashnikova and Yu.B. Yufryakov,
Phys.Atom.Nucl.60, (1997), 307-313
\bibitem{cm_light} P. Lacock {\it et al.}, UKQCD Collaboration, Phys. Rev. {\bf D}54, (1996), 6997-7009.
\bibitem{milc} C. Bernard {\it et. al.}, MILC Collaboration,
Nucl. Phys. {\bf B} (Proc. Suppl.) 53, (1997), 228-230.\\
C. Bernard {\it et. al.}, MILC Collaboration, hep-lat/9707008.
\bibitem{cm_and_peran} S. Perantonis and C. Michael, Nucl. Phys. {\bf B}347, (1990), 854-868. 
\bibitem{ford} I.J. Ford {\it et al.}, Phys. Lett. {\bf B}208, (1988), 286-290. 
\bibitem{nrqcd_improved} G.P. Lepage {\it et al.}, Phys. Rev. {\bf D}46, (1992), 4052-4067.
\bibitem{nrqcd_precision} C.T.H. Davies {\it et al.} Phys. Rev. {\bf
D}50, (1994), 6963-6977.
\bibitem{omv6} T. Manke {\it et al.}, UKQCD Collaboration, Phys. Lett. {\bf B}408, (1997), 308-314. 
\bibitem{trottier} H.D. Trottier, Phys. Rev. {\bf D}55, (1997), 6844-6851.
\bibitem{hybrid_constituent} A. Le Yaouanc {\it et al.}, Z. Phys. {\bf
C}28, (1985), 309-315.
\bibitem{griffiths} L.A. Griffiths {\it et al.}, Phys. Lett. {\bf
B}129, (1983), 351-356.
\bibitem{ron_hybrid} S. Catterall  {\it et al.}, UKQCD Collaboration, Phys. Lett. {\bf B}300, (1993), 393-399.
\bibitem{cm_fuzz} P. Lacock {\it et al.}, UKQCD Collaboration, Phys. Rev. {\bf D}51, (1995), 6403-6410.
\bibitem{jacobi_smearing}C.R. Allton {\it et al.}, UKQCD
Collaboration, Phys. Rev. {\bf D}47, (1993), 5128-5137.
\bibitem{numrec} W.H. Press {\it et al.}, ``Numerical recipes in Fortran'', (2nd Edition), Cambridge University Press (1992).
\bibitem{our_hybrid} T. Manke {\it et al.}, UKQCD Collaboration, to appear in
Nucl. Phys. {\bf B} (Proc. Suppl), proceedings of Lattice '97,
Edinburgh, hep-lat/9709001.
\bibitem{sara_hybrid} S. Collins {\it et al.}, UKQCD Collaboration, to
appear in Nucl. Phys. {\bf B} (Proc. Suppl), proceedings of Lattice '97,
Edinburgh, hep-lat/9710058.
\bibitem{pdb} R.M. Barnett {\it et al.}, Phys. Rev. {\bf D}54, (1996), 1.
\bibitem{close_break} F. E. Close and P. R. Page, Nucl. Phys. {\bf B}443, (1995), 233-254.

\bibitem{kuti_hybrid} K.J. Juge {\it et al.}, to appear in
Nucl. Phys. {\bf B} (Proc. Suppl), proceedings of Lattice '97,
Edinburgh, hep-lat/9709132.
\bibitem{colin_hybrid} K.J. Juge {\it et al.}, to appear in
Nucl. Phys. {\bf B} (Proc. Suppl), proceedings of Lattice '97,
Edinburgh, hep-lat/9709131.
\bibitem{flux_hybrid} T. Barnes {\it et al.}, Phys. Rev. {\bf D}52,
(1995), 5242-5256.
\end{thebibliography}
\end{document}